\documentclass[preprint,groupedaddress]{revtex4}

\usepackage{graphicx}
\usepackage{color}
\usepackage{amsfonts,amssymb}
\usepackage{natbib}

\begin{document}

\title{Turbulent transport of material particles :\\ An experimental study of finite size effects}

\author{Nauman M. Qureshi, Micka\"el Bourgoin, Christophe Baudet, Alain Cartellier, Yves Gagne}

\address{Laboratoire des \'Ecoulements G\'eophysiques et Industriels, CNRS/UJF/INPG UMR5519, BP53, 38041 Grenoble, France}

\begin{abstract}
We use an acoustic Lagrangian tracking technique, particularly adapted to measurements in open flows, and a versatile material particles generator (in the form of soap bubbles with adjustable size and density) to characterize Lagrangian statistics of finite sized, neutrally bouyant, particles transported in an isotropic turbulent flow of air. We vary the size of the particles in a range corresponding to turbulent inertial scales and explore how the turbulent forcing experienced by the particles depends on their size. We show that, while the global shape of the intermittent acceleration probability density function does not depend significantly on particle size, the acceleration variance of the particles decreases as they become larger in agreement with the classical $k^{-7/3}$ scaling for the spectrum of Eulerian pressure fluctuations in the carrier flow.  

\end{abstract}
\maketitle

Particle laden flows play an important role in several industrial applications (diesel engines, mixers, separative techniques, etc.), environment issues (atmospheric dispersion of pollutants, sediments transport, etc.) and many natural systems (plankton in seas, spores and pollen dispersion, water droplets in clouds, planetesimals in accretion disks, etc.). In most cases the carrier phase is a turbulent flow with eddies distributed over a wide range of relevant scales (called the inertial range of turbulence) : from the energy injection scale ($L$) down to the Kolmogorov dissipation scale ($\eta$). If the advected particles are neutrally buoyant and small (typically comparable in size or smaller than $\eta$) they behave as tracers for fluid particles. Such tracers are commonly used for single phase hydrodynamic measurements such as PIV, LDV and PTV. However, in many practical and natural situations, the transported particles have a density different from that of the surrounding fluid and/or a size larger than $\eta$. Their dynamics, which is then affected by particles inertia, deviates from that of fluid particles. The statistical description of inertial particles interacting with a turbulent flow remains a largely open question which clearly requires further experimental insight. From a theoretical point of view, even writing (not to mention solving) the particle motion equation in the most general case remains a challenge and only limit cases have been treated at present. For instance, in most analytical approaches, particles are assumed to be small compared to the scales of turbulence, so that the fluid around is assumed locally laminar. In this framework, particles are approximated by point-particles and the forces acting on them are : buoyancy, acceleration of the undisturbed flow, the Stokes drag, the added mass and the Basset-Boussinesq history term \cite{bib:maxey1983}. In this limit an approximate equation of motion can be written and coupled to turbulence models or numerical simulations to predict the motion of the advected inertial particles \cite{bib:maxey1983, bib:bec2005_JFM, bib:balkovsky2001_PRL, bib:Zaichik2004}. This approach for instance predicts the wellknown preferential concentration and enhancement of settling velocity of heavy particles by the turbulence. Such recent numerical studies \cite{bib:bec2005_JFM} have also shown to correctly describe acceleration statistics measured for small water droplets (with sub-Kolmogorov size) transported in a turbulent air flow \cite{bib:ayya2006}. However when the particle size is of the order of Kolmogorov scale or larger, such point-particle approximations might become inapropriate. 

In this letter, we present an experimental study of the role played by particles finite size on their Lagrangian small scale dynamics as they are advected by the turbulence. In order to decouple finite size effects from density and seeding concentration effects, we only consider here neutrally buoyant particles (the case of heavy particles will be addressed in a forthcoming work) with an extremely low seeding density. Particles are rigid (we use soap bubbles, with Weber number much smaller than one in experimental conditions), spherical and characterized only by the ratio $\phi$ of their diameter $D$ to the Kolmogorov scale of the carrier flow, $\phi=D/\eta$. We vary the particles diameter over a range corresponding to inertial scales ($\phi > 1$).

\begin{figure}[b]
\centering
\includegraphics[height=2.7cm]{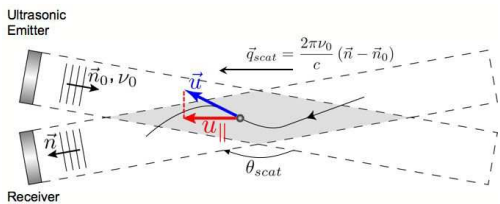}\\
\caption{Principle of acoustical Doppler velocimetry : an ultrasonic plane wave of frequency $\nu_0$ is emitted in the direction $\vec{n}_0$ by a transducer. A receiver records the wave scattered in the direction $\vec{n}$ by a moving particle with velocity $\vec{u}$. The component $u_{//}$ of the particle velocity along the scattering vector $\vec{q}_{scat}$ is directly measured by the Doppler shift of the frequency of the recorded scattered wave.}
\label{fig:1}
\end{figure}
An intuitive phenomenology for finite size effects on the small scales motion of  finite sized particles advected by a turbulent flow relies on the idea that such particles are not sensitive to flow gradients at scales smaller than their own size. In this phenomenology, the turbulent energy spectrum of spatial scales actually forcing the particles dynamics is therefore somehow filtered, as only turbulent structures at scales typically larger than $D$ do contribute to the advection of the particles. We address here the question of the relevance of this phenomenology and of how the Langrangian dynamics of the particles (caracterized mostly by their acceleration) is affected by such spatial filtering of the turbulent field due to the particles finite size.


Our experiment runs in a large closed wind tunnel with a measurement section of $0.75~\textrm{m} \times 0.75~\textrm{m}$ where the turbulence is generated behind a grid with a mesh size of 6~cm and reproduces almost ideal isotropic turbulence. The results reported here were obtained with a mean velocity of the fluid $U=15\;\textrm{m}\cdot\rm{s}^{-1}$ and a turbulence level $u_{rms}/U \simeq 3\%$. The corresponding Reynolds number, based on Taylor microscale, is of the order of $R_\lambda=160$. The dissipation scale $\eta$ is 240~$\mu$m and the energy injection scale $L$ is 6~cm. As particles we use soap bubbles inflated with helium in order to adjust the density to neutrally buoyant in air \cite{bib:poulain2004}. The size of the bubbles can be adjusted between 1.5~mm and 6~mm, corresponding to ratios to Kolmogorov scale $\phi$ ranging from 7 to 25. For a given setting of the bubble size, the standard deviation of the size distribution is relatively low (typically of the order of $D_{rms}\sim150~\mu$m), except for the smallest bubbles ($D\sim1.5$~mm) for which the nozzle generating bubbles has to be pushed to an extreme regime where it becomes slightly unstable and produces a wider distribution of particles diameter (typically $D_{rms}\sim500~\mu$m in this regime). All the different particles we have tested have a diameter which corresponds to the carrier turbulence inertial range scales. The smallest particles ($\phi\sim7$) approach the dissipation scale while the largest ones ($\phi\sim25$) approach the injection scale. The seeding density is extremely low (particles are injected individually), so that no global modulation of the turbulence by the particles is expected. Particles are individually tracked by 1D Lagrangian acoustic Doppler velocimetry \cite{bib:gervais2007_ExpFluids,bib:mordant2005_RSI} (figure \ref{fig:1}). We measure the longitudinal (streamwise) velocity component $v_z$ of the particles along their trajectory. As they travel accross the measurement volume, the particles can be tracked during approximately 50~ms at most, which corresponds to several dissipation time scales $\tau_\eta=(\nu/\epsilon)^{1/2}\simeq 3.8$~ms ($\nu=1.5 \cdot 10^{-5} \; \textrm{m} ^2 \cdot \textrm{s} ^{-1} $ is the air viscosity at working temperature and $\epsilon\simeq1.0 \; \textrm{m} ^2 \cdot \textrm{s}^{-3}$ the average energy dissipation rate per unit mass at the measurement volume location). For each set of particle sizes, we record approximately 4000 tracks at a samplimg rate of 32768~Hz, giving more than $10^6$ data points for each set.

\begin{figure}[h!]
\begin{minipage}{\linewidth}
\centering
\includegraphics[width=6cm]{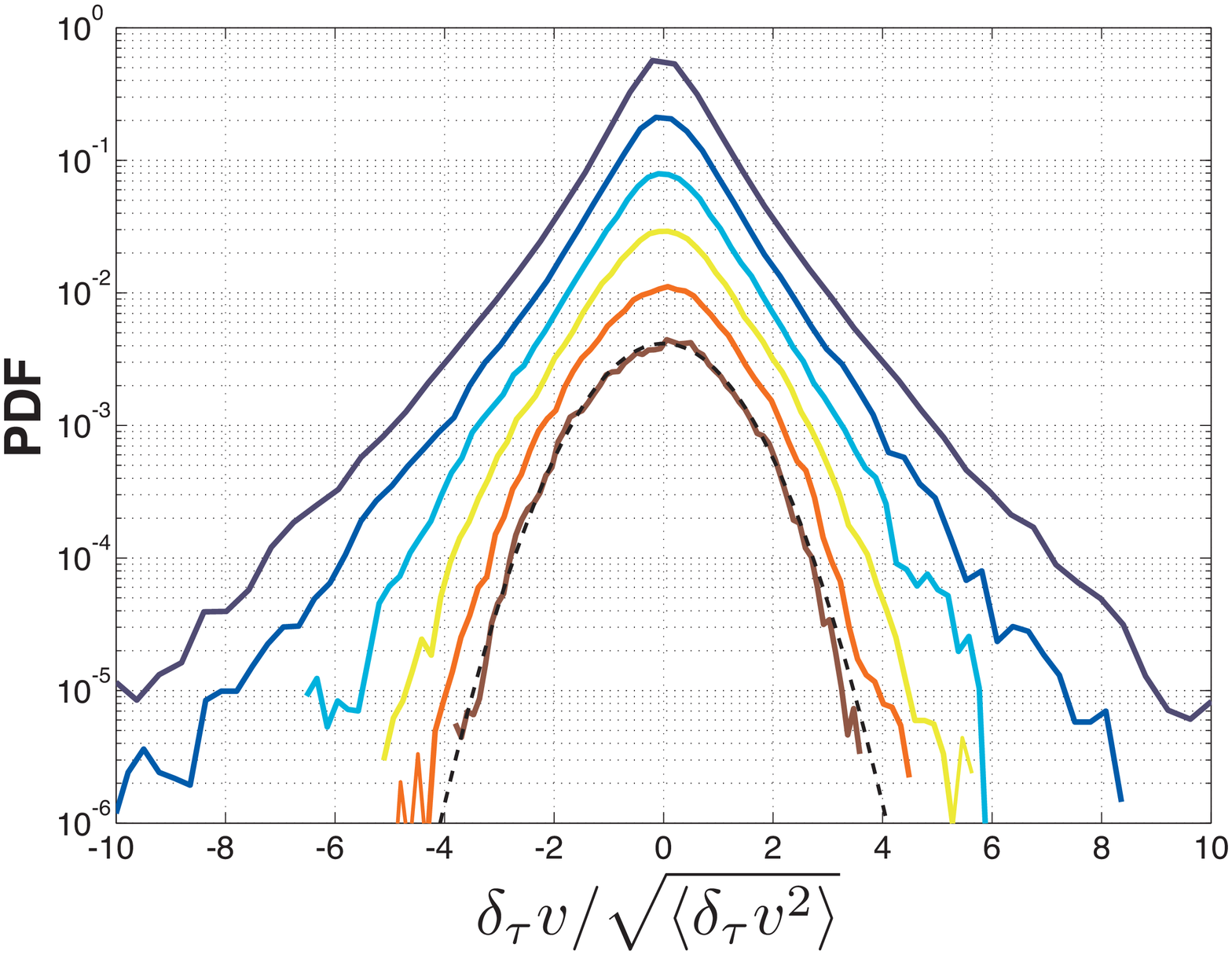}
\end{minipage}
\begin{minipage}{\linewidth}
\includegraphics[width=6cm]{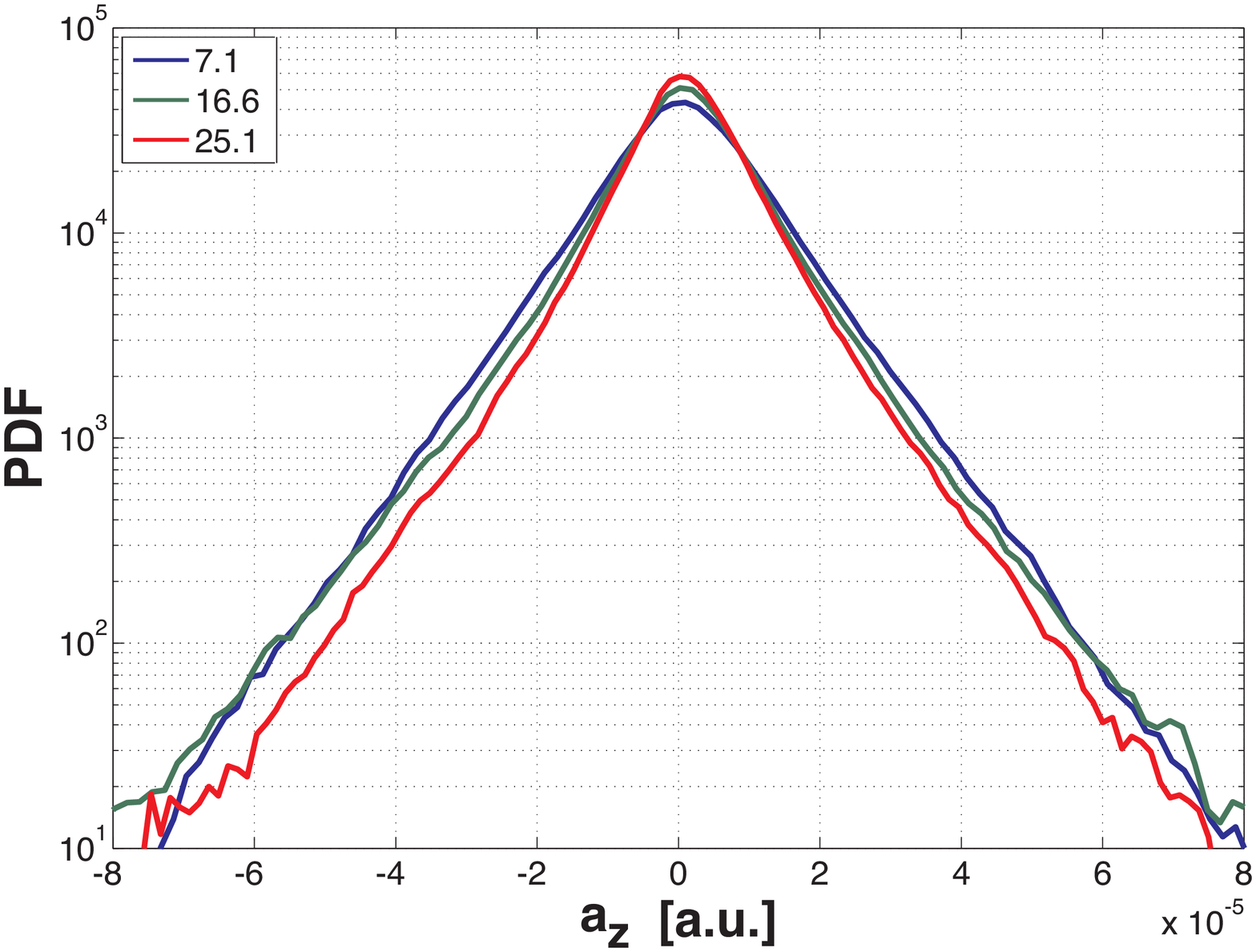}
\end{minipage}
\begin{minipage}{\linewidth}
\includegraphics[width=6cm]{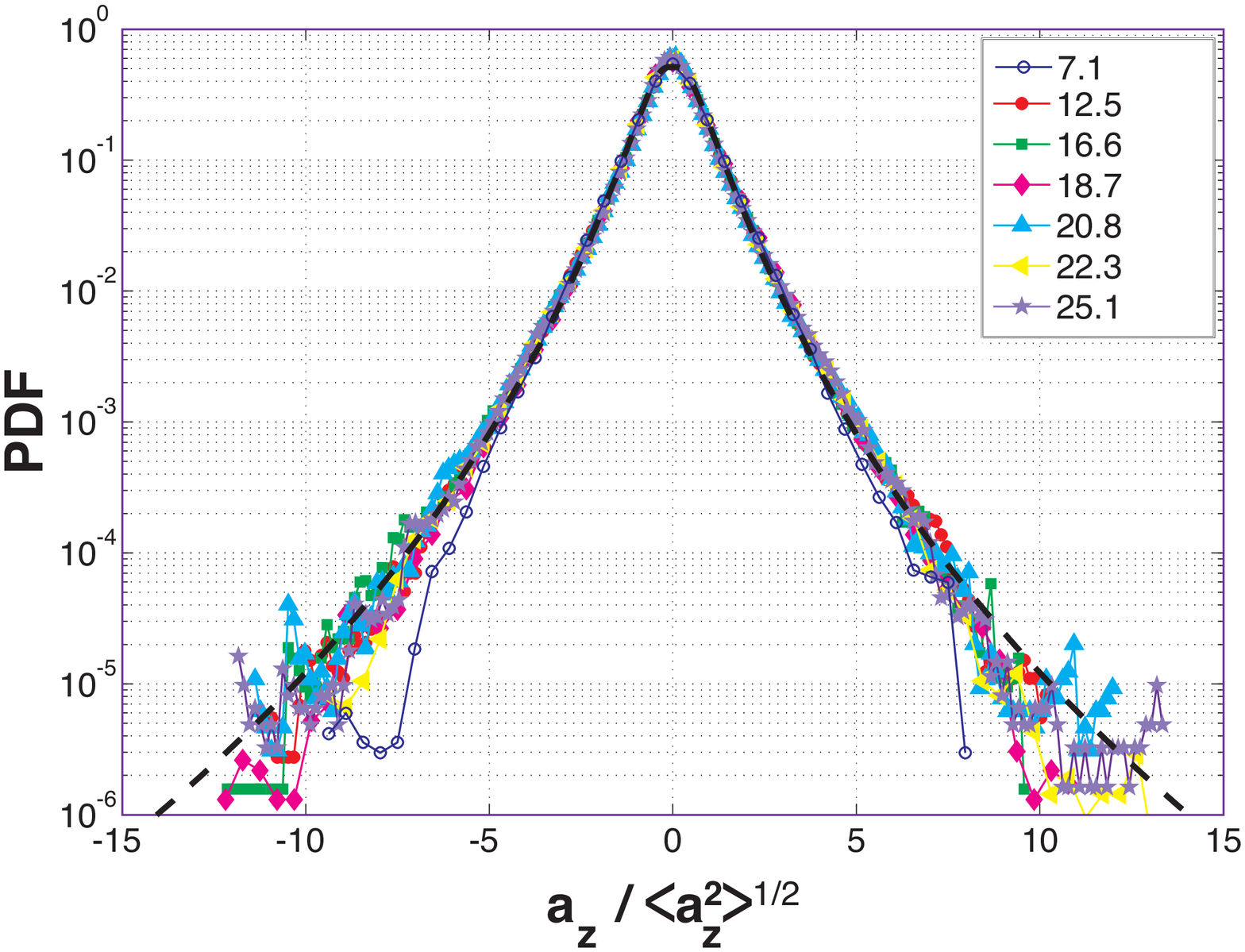}
\end{minipage}
\caption{(a) Lagrangian velocity increments probability density functions (PDF) for 6~mm bubbles ($\phi \sim 25.1$) for different time lags $\tau = \tau_\eta/50,\, 2\tau_\eta,\, 4\tau_\eta,\, 6\tau_\eta,\, 8\tau_\eta \rm{ and } 10\tau_\eta$ from top to bottom ; curves have been shifted vertically for readability ; the dashed line represents a gaussian distribution of variance 1. (b) Non normalized acceleration PDF for 3 different sizes of particles ($\phi=7.1\,;\,16.6 \, {\rm{ and }} \, 25.1$). (c) Collapse of the acceleration normalized (to variance 1) PDFs for different particles sizes ($\phi = 7.1 \rightarrow 25.1$) ; the dashed line represents a fit by relation (\ref{eq:AccComp}) with $s=0.62$. }\label{fig:Increments}
\end{figure}

In order to characterize the particles dynamics at different time-scales, we analyze the Lagrangian velocity increments $\delta_\tau v_z(t)=v_z(t+\tau)-v_z(t)$. As an example, figure \ref{fig:Increments}a represents the probability density functions (PDF) of the Lagrangian velocity increments normalized by their standard deviation for increasing values of $\tau$ for the largest particles ($\phi=25$, $D=6~$mm). The continuous deformation of the PDF from gaussian at large time-scales to the development of stretched exponential tails at dissipative time-scales, is the signature of an intermittent Lagrangian dynamics \cite{bib:mordant2001,bib:chevillard2003_PRL}, which we found here to remain present even for particles approaching in size the energy injection scale. The wide stretched exponential tails of the PDF for the smallest time lags $\tau$ show that large particles still experience very high acceleration events with a probability much higher than gaussian, as previously reported for small tracers \cite{bib:LaPorta2002}. Here after, we focus our analysis on the acceleration statistics of the particles (which is of particular interest since it directly reflects the actual forcing particles experience from the turbulent carrier flow) and on how the turbulent forcing acting on a particle depends on its size. Technically, acceleration is obtained by convolution of the Doppler velocity signal with a differentiated gaussian kernel. This procedure simultaneously differentiates and filters the signal from unavoidable experimental noise \cite{bib:mordant2004_PhysicaD}.
 
Figure \ref{fig:Increments}b represents the acceleration PDF (here acceleration has not been normalized by its standard deviation) for three different particle sizes. Acceleration PDF clearly narrows and peaks around zero as the particle dimension increases. This gives a clear evidence that depending on their size, particles experience a turbulent forcing with different statistical signature. Larger particles present a narrower acceleration distribution than smaller, meaning that they are less sensitive to extreme turbulent events. An interesting observation is that the evolution of the PDF with the particle diameter is mostly associated solely to a decrease of the acceleration variance of the particles as their diameter increases. Indeed, when we normalize the acceleration component $a_z$ of the particles by its standard deviation $a_{z,\rm rms}$, the corresponding PDFs are found to collapse for all particle diameters (figure \ref{fig:Increments}c). Therefore, the global shape of the acceleration PDF, normalized to variance 1, is found to be preserved for all particle sizes. (We note that the PDF for the smallest particles ($\phi=7.1$) does not collapse as well as the others, especially the PDF tails deviate. This is very likely due to the larger dispersion in the size distribution of the smallest particles which tends to overstimate the actual acceleration standard deviation and to alter the global shape of the PDF.) Over the range $\left|a_z\right| < 10 \; a_{z,\rm rms}$ the shape of the normalized PDF is well described by the function :

\begin{equation}\label{eq:AccComp}
{\cal {P}}(x)=\frac{e^{3s^2/2}}{4\sqrt{3}} \left[ 1- {\rm erf} \left(\frac{ {\rm ln}\left(\left| x / \sqrt{3} \right| \right)+2s^2} {\sqrt{2}s} \right) \right]
\end{equation}

\noindent where $s$ is a parameter related to the global shape of the PDF. 
This form for the PDF of an acceleration component was previously suggested for the case of fluid tracer particles (with size comparable to $\eta$) \cite{bib:mordant2004b} and was shown to be related to a lognormal distribution (of variance $s^2$) of the whole vector acceleration magnitude \cite{bib:moninbookII} under the condition of isotropic turbulence. We found here that this form still correctly describes the acceleration component statistics of finite sized particles and considering the good isotropic conditions in our experiment, we can expect acceleration magnitude of finite sized particles to be indeed well described by a lognormal distribution. Our experimental data is best fitted with a value for the parameter $s$ in relation (\ref{eq:AccComp}) equal to 0.62, which corresponds to a flatness for the acceleration fluctuations equal to 8.4 (in good agreement with the direct estimation of the flatness from the data).
This value is significantly smaller than values previously reported for fluid tracers (of the order of 20) at similar Reynolds number  \cite{bib:voth2002,bib:mordant2001}, although a unambiguous comparison between those experiments and ours is made difficult due experimental specificities ; in particular while our grid generated turbulence is largely isotropic, the von K\'arm\'an flow studied in \cite{bib:voth2002,bib:mordant2001} has a large scale anisotropy, which have been shown to possibly affect small turbulent scales, even in high Reynolds numbers regimes \cite{bib:Ouellette2006_NJP}. 
If, in spite of these differences, a continuity between those previous experiments and ours is however attempted, the difference between the acceleration flatness between small fluid tracers and our finitie sized particles would mean that a rapid decrease of the acceleration flatness must occur for particles sizes around a few $\eta$ ($\phi \sim {\cal{O}}(1)$) as the size increases from dissipative to inertial range scales. In any case, our results show that acceleration statistics reach an asymptotic regime for inertial range sizes, regardless of the particles size (at least up to $\phi \sim 30$), described by the PDF in figure \ref{fig:Increments}c and equation (\ref{eq:AccComp}). An interesting and new result to be stressed is that the same identical non-gaussian acceleration PDF is preserved even for particles with size approaching injection scales. 
Based on the classical image of Eulerian intermittency of the carrier turbulent velocity field (where velocity increments are non-gaussian at small scales and tend continuously toward a gaussian at larger scales), one might have accordingly expected a monotical trend of particles acceleration statistics toward a gaussian as particles size approaches large scales. Similarly, one might have expected the acceleration statistics of particles of size $D$ to be somehow related to the Lagrangian velocity increments statistics  calculated along fluid tracers trajectories at a time lag $\tau_D$ given by the eddy turnover time at scale $D$, which because of Lagrangian velocity intermittency \cite{bib:mordant2001,bib:chevillard2003_PRL}, would also tend continuously toward a gaussian at large scales. This not being the case, we can conclude that the turbulent forcing of finite sized particles with inertial range sizes is not trivially related to intermittency (neither Eulerian nor Lagrangian) of the turbulent velocity field around the particle; in particular acceleration PDF for particles with a given size cannot be simply deduced from statistics of velocity increments at scales corresponding to the particle size.

A possible reason to this is that from a dynamical point of vue, acceleration reflects the forcing felt by the particle, which, for inertial range scales, is mostly due to turbulent fluctuations of the pressure gradients in the vicinity of the particule. Under an ergodic assumption, particles acceleration statistics can therefore be related to Eulerian pressure increments statistics, which are indeed not trivially related to velocity increments statistics \cite{bib:hill1993_BullAmPhys}. In particular, in this context the acceleration variance of a particle with diameter $D$ is related to the second moment of the pressure increments at the scale of the particle: 
\begin{equation}\label{eq:acc}
\left<a_z^2\right>_{\rm{particle}}(D) \propto \frac{S^P_2(D)}{D^2},
\end{equation}
where $S^P_2(r)=\left< \left(p(\vec{x}+\vec{r})-p(\vec{x})\right)^2\right>$, is the Eulerian second order pressure structure function (which only depends on $r=|\vec{r}|$ in homogeneous isotropic conditions). In the frame of Kolmogorov 41 phenomenology, the classical inertial scaling for $S^P_2(r)$ is $(\epsilon r)^{4/3}$ \cite{bib:obukhov1949,bib:moninbookII} (or equivalently $\epsilon^{4/3}k^{-7/3}$ for the pressure spectrum in $k$ wave number space). Although this scaling is still controversial \cite{bib:vedula1999_PoF,bib:gotoh2001_PRL}, there are experimental \cite{bib:george1984_JFM,bib:nelkin1994_AdvPhys} and numerical \cite{bib:vedula1999_PoF} evidences suggesting that it is most likely correct in ranges of Reynolds numbers and scales consistent with our flow and particle sizes. As a consequence, the acceleration variance of particles with diameter $D$ in inertial range scales should follow the scaling:

\begin{equation}\label{eq:modifiedHY}
\left<a_z^2\right>_{\rm{particle}}(D) = a'_0 \epsilon^{4/3}D^{-2/3}.
\end{equation}

\begin{figure}[t]
\centering
\includegraphics[width=7cm]{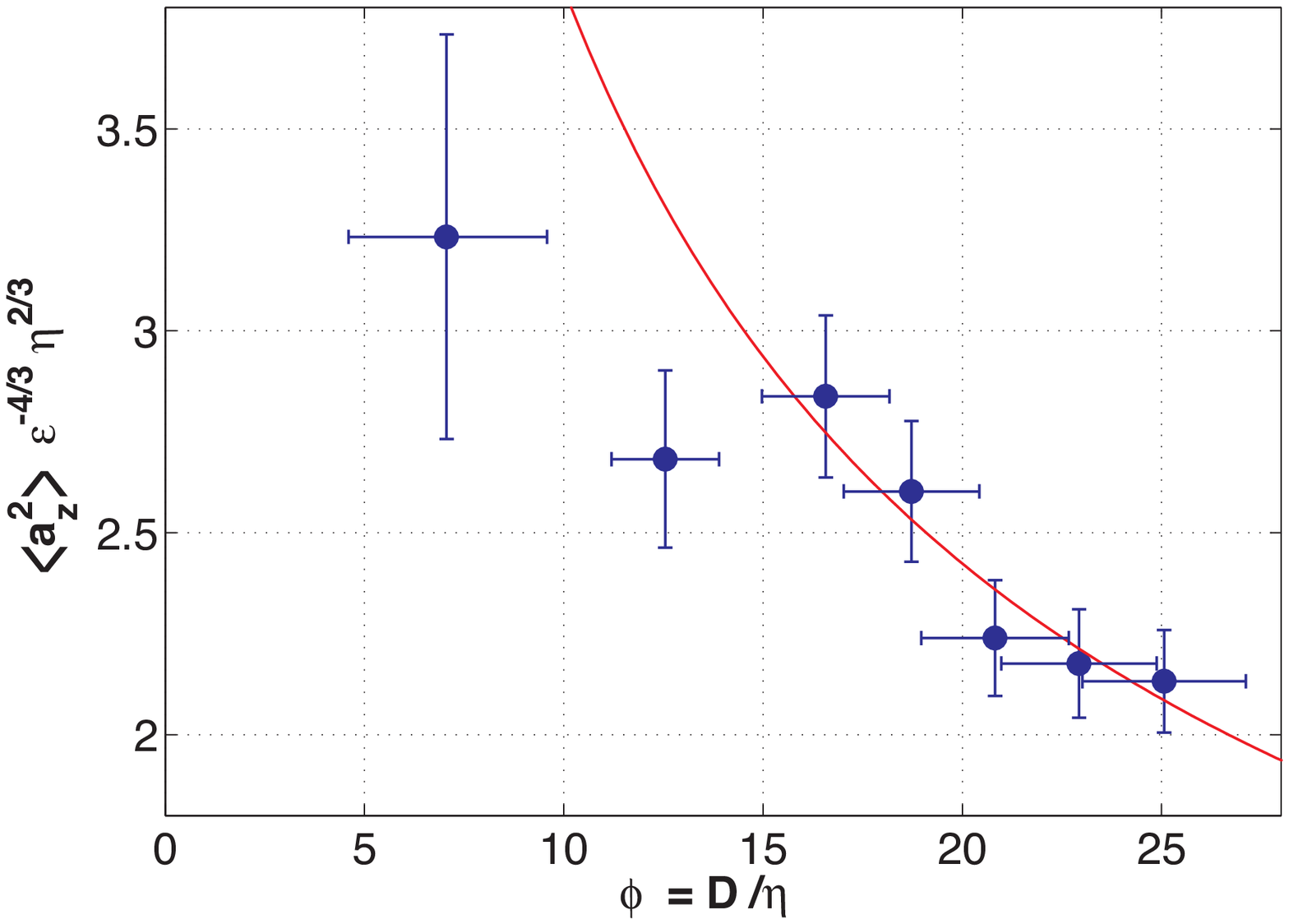}
\includegraphics[width=7cm]{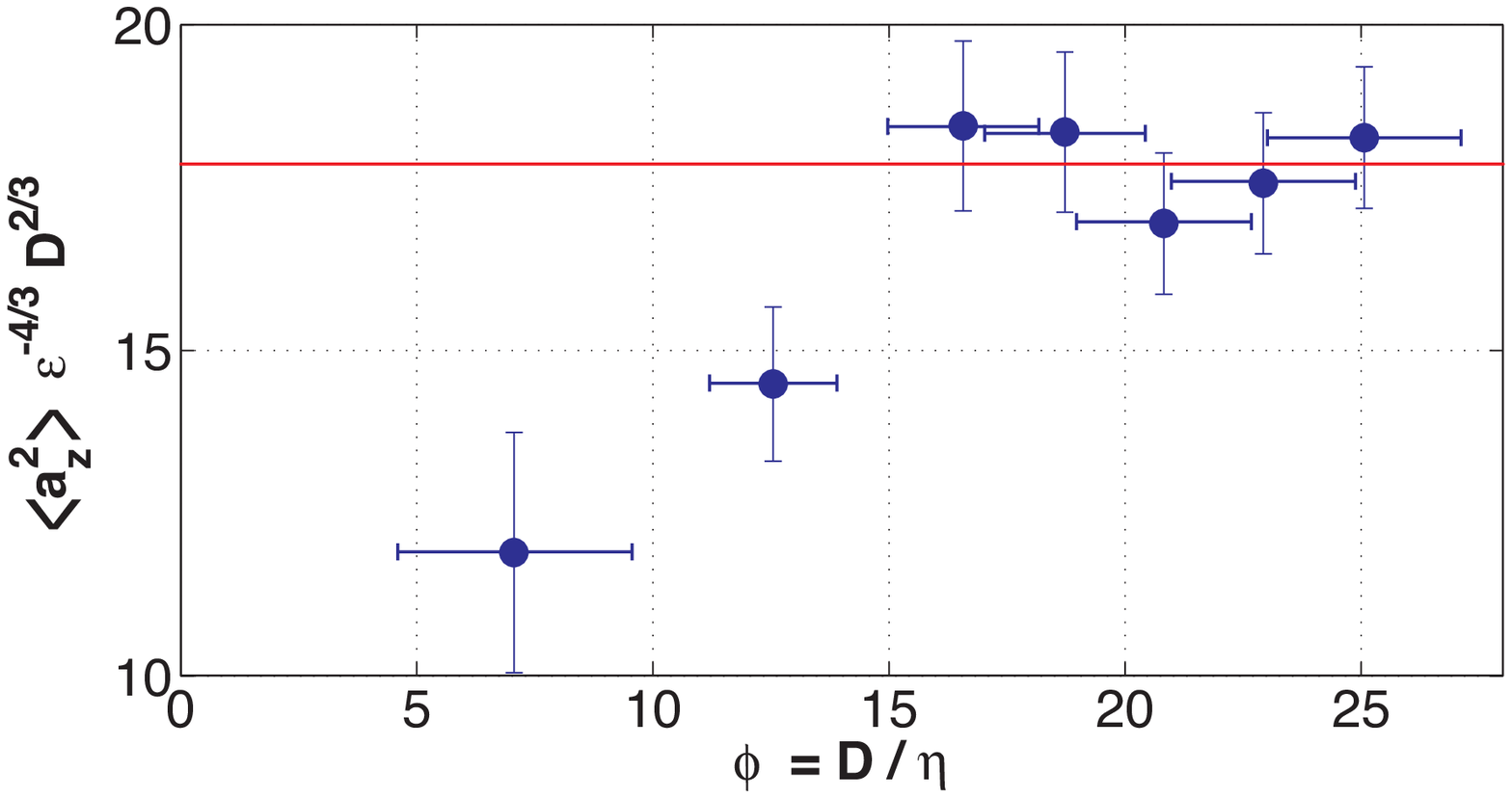}
\caption{(a) Dimensionless acceleration variance as a function of particle size ; errorbars are mostly associated to the particle size distribution and to the uncertainty on the determination of $\epsilon$. The point for the smallest value of $\phi$ has larger errorbars due bubble nozzle instability which produces a larger size distribution in this regime, tending very likely to overestimate particles acceleration variance.(b) Compensated acceleration variance as a function of particle size. In both figures the solid line represents a $D^{-2/3}$ law.}\label{fig:Arms}
\end{figure}

\noindent This scaling can also be directly derived by simple dimensional considerations in the framework of Kolmogorov's inertial range hypotheses \cite{bib:moninbookII}, but we stress here the dynamical correspondance with the pressure increments statistics. Relation (\ref{eq:modifiedHY}) predicts a decrease of the acceleration variance with increasing particles diameter. Our measurements show indeed that while the global shape of the normalized acceleration PDF is found not to depend on particle size, a 40\% decrease is measured for the acceleration variance itself between the smallest and the largest particles we have tested. Figure \ref{fig:Arms}a shows more precisely the decrease of the acceleration variance (made dimensionless by $ \epsilon^{4/3} \eta^{-2/3}$) with the particle size. Errorbars are mostly associated to the particles size distribution and to uncertainties in the determination of $\epsilon$. The decay of acceleration variance is found in good agreement with relation (\ref{eq:modifiedHY}) for particles larger than about 4~mm ($\phi > 15$), as it can be more clearly seen on the compensated plot on figure \ref{fig:Arms}b, where an asymptotic plateau (around $a'_0 \sim 18$) is reached for the largest particles. 

In the limit of very small particles (ideally $\phi \ll 1$) 
acceleration variance should saturate to an asymptotic value corresponding to the intrinsinc acceleration variance of the turbulent flow itself. In this limit, scaling (\ref{eq:modifiedHY}) which has been derived from inertial range considerations, should be replaced by its equivalent small scale (dissipative) form, given by the usual Heiseinberg-Yaglom relation for fluid particles \cite{bib:moninbookII}:

\begin{equation}\label{eq:HY}
\left<a_z^2\right>_{\rm{fluid}}= a_0 \epsilon^{4/3} \eta^{-2/3}.
\end{equation}

\noindent Using high speed optical particle tracking Voth and collaborators \cite{bib:voth2002} have shown that acceleration variance of particles such that $\phi <  5$  is already correctly described by the dissipative scaling (\ref{eq:HY}) with a Reynolds number dependent constant $a_0$ (reasonably, the dissipative scaling is reached far before the ideal limit $\phi \ll 1$). Figure \ref{fig:Arms}b shows a clear deviation from the inertial scaling (\ref{eq:modifiedHY}) as soon as $\phi < 15$. Therefore, we can conclude that the crossover between the inertial scaling (\ref{eq:modifiedHY}) and the dissipative scaling (\ref{eq:HY}) for the acceleration variance occurs for particles with size in the range $5 < \phi < 15$. Although we are not able here to study particles smaller than $\phi < 7$ (for technical reasons related to our bubble generator nozzle), an extrapolation to the limit $\phi \rightarrow 0$ from figure \ref{fig:Arms}a gives a rough estimate for the constant $a_0$ in (\ref{eq:HY}) at the working Reynolds number ($R_\lambda \simeq 160$) of the order of $a_0\sim3$ which is in good agreement with  the experimental value by Voth and collaborators \cite{bib:voth2002} and from earlier predictions from direct numerical simulations \cite{bib:vedula1999_PoF,bib:gotoh2001_PRL}.

We have studied Lagrangian statistics of finite sized neutrally buoyant particles transported in a trubulent flow of air, with a particular focus on particles acceleration. These are the first systematic experimental measurements of finite size effects on the transport of material particles ever done in an open and nearly isotropic turbulent flow ; they were made possible by  the combination of a novel acoustical Lagrangian tracking system and a versatile particle generator. We have shown that particle dynamics displays intermittency and that particle acceleration statistics is described by the same probability density function for all particles with inertial range sizes, where only the acceleration variance displays a significant dependence on particle size. In particular acceleration fluctuations remain non-gaussian even for large particles (approaching the injection scale). The decrease of the acceleration variance with particle size is in good agreement with classical inertial scaling for pressure increments in the carrier flow, what supports the connection between the Lagrangian dynamics of the particles and Eulerian statistics of the carrier turbulent flow via the pressure increments field. A deeper theoretical and numerical insight on pressure increments statistics at different spatial scales (for instance in terms of a systematic analysis of pressure increments PDF) would very likely help the further understanding and the modeling of such finite size effects.
Our results give an experimental support to the idea, generally assumed for the modelling of particles transport, that turbulent pressure fluctuations only at scales larger than the particles size do contribute to the transport and the dispersion of the particles, which is of particular interest in the frame of Large Eddy Simulations (LES) for the study of particles-turbulence interactions. Finally, in a purely hydrodynamics context, the present experiments also open new possibilities for characterizing Eulerian statistics of turbulent pressure fluctuations at inertial scales by means of Lagrangian measurements of the acceleration of finite sized particles.
\begin{acknowledgments}
We deeply acknowledge Joseph Virone for the technical support in manufacturing parts of the experimental setup. We also deeply acknowledge Nicolas Mordant, Jean-Fran\c{c}ois Pinton and Romain Volk for fruitful discussions.
\end{acknowledgments}

\bibliographystyle{apsrev}
\bibliography{main}

\begin{thebibliography}{22}
\expandafter\ifx\csname natexlab\endcsname\relax\def\natexlab#1{#1}\fi
\expandafter\ifx\csname bibnamefont\endcsname\relax
  \def\bibnamefont#1{#1}\fi
\expandafter\ifx\csname bibfnamefont\endcsname\relax
  \def\bibfnamefont#1{#1}\fi
\expandafter\ifx\csname citenamefont\endcsname\relax
  \def\citenamefont#1{#1}\fi
\expandafter\ifx\csname url\endcsname\relax
  \def\url#1{\texttt{#1}}\fi
\expandafter\ifx\csname urlprefix\endcsname\relax\def\urlprefix{URL }\fi
\providecommand{\bibinfo}[2]{#2}
\providecommand{\eprint}[2][]{\url{#2}}

\bibitem[{\citenamefont{Maxey and Riley}(1983)}]{bib:maxey1983}
\bibinfo{author}{\bibfnamefont{M.~R.} \bibnamefont{Maxey}} \bibnamefont{and}
  \bibinfo{author}{\bibfnamefont{J.~J.} \bibnamefont{Riley}},
  \bibinfo{journal}{Physics of Fluids} \textbf{\bibinfo{volume}{26}},
  \bibinfo{pages}{883} (\bibinfo{year}{1983}).

\bibitem[{\citenamefont{Balkovsky et~al.}(2001)\citenamefont{Balkovsky,
  Falkovich, and Fouxon}}]{bib:balkovsky2001_PRL}
\bibinfo{author}{\bibfnamefont{E.}~\bibnamefont{Balkovsky}},
  \bibinfo{author}{\bibfnamefont{G.}~\bibnamefont{Falkovich}},
  \bibnamefont{and} \bibinfo{author}{\bibfnamefont{A.}~\bibnamefont{Fouxon}},
  \bibinfo{journal}{Physical Review Letters} \textbf{\bibinfo{volume}{86}},
  \bibinfo{pages}{2790} (\bibinfo{year}{2001}).

\bibitem[{\citenamefont{Bec et~al.}(2005)\citenamefont{Bec, Biferale, Boffetta,
  Celani, Cencini, Lanotte, Musacchio, and Toschi}}]{bib:bec2005_JFM}
\bibinfo{author}{\bibfnamefont{J.}~\bibnamefont{Bec}},
  \bibinfo{author}{\bibfnamefont{L.}~\bibnamefont{Biferale}},
  \bibinfo{author}{\bibfnamefont{G.}~\bibnamefont{Boffetta}},
  \bibinfo{author}{\bibfnamefont{A.}~\bibnamefont{Celani}},
  \bibinfo{author}{\bibfnamefont{M.}~\bibnamefont{Cencini}},
  \bibinfo{author}{\bibfnamefont{A.}~\bibnamefont{Lanotte}},
  \bibinfo{author}{\bibfnamefont{S.}~\bibnamefont{Musacchio}},
  \bibnamefont{and} \bibinfo{author}{\bibfnamefont{F.}~\bibnamefont{Toschi}},
  \bibinfo{journal}{Journal of Fluid Mechanics}  (\bibinfo{year}{2005}).

\bibitem[{\citenamefont{Zaichik and Alipchenkov}(2004)}]{bib:Zaichik2004}
\bibinfo{author}{\bibfnamefont{L.}~\bibnamefont{Zaichik}} \bibnamefont{and}
  \bibinfo{author}{\bibfnamefont{V.~M.} \bibnamefont{Alipchenkov}},
  \bibinfo{journal}{High Temperature} \textbf{\bibinfo{volume}{42}},
  \bibinfo{pages}{426} (\bibinfo{year}{2004}).

\bibitem[{\citenamefont{Ayyalasomayajula
  et~al.}(2006)\citenamefont{Ayyalasomayajula, Gylfason, Collins, Bodenschatz,
  and Warhaft}}]{bib:ayya2006}
\bibinfo{author}{\bibfnamefont{S.}~\bibnamefont{Ayyalasomayajula}},
  \bibinfo{author}{\bibfnamefont{A.}~\bibnamefont{Gylfason}},
  \bibinfo{author}{\bibfnamefont{L.}~\bibnamefont{Collins}},
  \bibinfo{author}{\bibfnamefont{E.}~\bibnamefont{Bodenschatz}},
  \bibnamefont{and} \bibinfo{author}{\bibfnamefont{Z.}~\bibnamefont{Warhaft}},
  \bibinfo{journal}{Physical Review Letters} \textbf{\bibinfo{volume}{97}}
  (\bibinfo{year}{2006}).

\bibitem[{\citenamefont{Poulain et~al.}(2004)\citenamefont{Poulain, Mazellier,
  Gervais, Gagne, and Baudet}}]{bib:poulain2004}
\bibinfo{author}{\bibfnamefont{C.}~\bibnamefont{Poulain}},
  \bibinfo{author}{\bibfnamefont{N.}~\bibnamefont{Mazellier}},
  \bibinfo{author}{\bibfnamefont{P.}~\bibnamefont{Gervais}},
  \bibinfo{author}{\bibfnamefont{Y.}~\bibnamefont{Gagne}}, \bibnamefont{and}
  \bibinfo{author}{\bibfnamefont{C.}~\bibnamefont{Baudet}},
  \bibinfo{journal}{Flow, Turbulence and Combustion}
  \textbf{\bibinfo{volume}{72}}, \bibinfo{pages}{245} (\bibinfo{year}{2004}).

\bibitem[{\citenamefont{Gervais et~al.}(2007)\citenamefont{Gervais, Baudet, and
  Gagne}}]{bib:gervais2007_ExpFluids}
\bibinfo{author}{\bibfnamefont{P.}~\bibnamefont{Gervais}},
  \bibinfo{author}{\bibfnamefont{C.}~\bibnamefont{Baudet}}, \bibnamefont{and}
  \bibinfo{author}{\bibfnamefont{Y.}~\bibnamefont{Gagne}},
  \bibinfo{journal}{Experiments in Fluids} \textbf{\bibinfo{volume}{42}},
  \bibinfo{pages}{371} (\bibinfo{year}{2007}), ISSN \bibinfo{issn}{0723-4864}.

\bibitem[{\citenamefont{Mordant et~al.}(2005)\citenamefont{Mordant, Metz,
  Pinton, and Michel}}]{bib:mordant2005_RSI}
\bibinfo{author}{\bibfnamefont{N.}~\bibnamefont{Mordant}},
  \bibinfo{author}{\bibfnamefont{P.}~\bibnamefont{Metz}},
  \bibinfo{author}{\bibfnamefont{J.~F.} \bibnamefont{Pinton}},
  \bibnamefont{and} \bibinfo{author}{\bibfnamefont{O.}~\bibnamefont{Michel}},
  \bibinfo{journal}{REVIEW OF SCIENTIFIC INSTRUMENTS}
  \textbf{\bibinfo{volume}{76}}, \bibinfo{pages}{025105}
  (\bibinfo{year}{2005}), ISSN \bibinfo{issn}{0034-6748}.

\bibitem[{\citenamefont{Mordant et~al.}(2001)\citenamefont{Mordant, Metz,
  Michel, and Pinton}}]{bib:mordant2001}
\bibinfo{author}{\bibfnamefont{N.}~\bibnamefont{Mordant}},
  \bibinfo{author}{\bibfnamefont{P.}~\bibnamefont{Metz}},
  \bibinfo{author}{\bibfnamefont{O.}~\bibnamefont{Michel}}, \bibnamefont{and}
  \bibinfo{author}{\bibfnamefont{J.-F.} \bibnamefont{Pinton}},
  \bibinfo{journal}{Physical Review Letters} \textbf{\bibinfo{volume}{87}},
  \bibinfo{pages}{214501} (\bibinfo{year}{2001}).

\bibitem[{\citenamefont{Chevillard et~al.}(2003)\citenamefont{Chevillard, Roux,
  L{\'e}v{\^e}que, Mordant, Pinton, and Arn{\'e}odo}}]{bib:chevillard2003_PRL}
\bibinfo{author}{\bibfnamefont{L.}~\bibnamefont{Chevillard}},
  \bibinfo{author}{\bibfnamefont{S.~G.} \bibnamefont{Roux}},
  \bibinfo{author}{\bibfnamefont{{\'E}.}~\bibnamefont{L{\'e}v{\^e}que}},
  \bibinfo{author}{\bibfnamefont{N.}~\bibnamefont{Mordant}},
  \bibinfo{author}{\bibfnamefont{J.-F.} \bibnamefont{Pinton}},
  \bibnamefont{and}
  \bibinfo{author}{\bibfnamefont{A.}~\bibnamefont{Arn{\'e}odo}},
  \bibinfo{journal}{Physical Review Letters} \textbf{\bibinfo{volume}{91}},
  \bibinfo{pages}{214502} (\bibinfo{year}{2003}).

\bibitem[{\citenamefont{LaPorta et~al.}(2002)\citenamefont{LaPorta, Voth,
  Crawford, Alexander, and Bodenschatz}}]{bib:LaPorta2002}
\bibinfo{author}{\bibfnamefont{A.}~\bibnamefont{LaPorta}},
  \bibinfo{author}{\bibfnamefont{G.~A.} \bibnamefont{Voth}},
  \bibinfo{author}{\bibfnamefont{A.~M.} \bibnamefont{Crawford}},
  \bibinfo{author}{\bibfnamefont{J.}~\bibnamefont{Alexander}},
  \bibnamefont{and}
  \bibinfo{author}{\bibfnamefont{E.}~\bibnamefont{Bodenschatz}},
  \bibinfo{journal}{Nature} \textbf{\bibinfo{volume}{409}},
  \bibinfo{pages}{1017} (\bibinfo{year}{2002}).

\bibitem[{\citenamefont{Mordant
  et~al.}(2004{\natexlab{a}})\citenamefont{Mordant, Crawford, and
  Bodenschatz}}]{bib:mordant2004_PhysicaD}
\bibinfo{author}{\bibfnamefont{N.}~\bibnamefont{Mordant}},
  \bibinfo{author}{\bibfnamefont{A.~M.} \bibnamefont{Crawford}},
  \bibnamefont{and}
  \bibinfo{author}{\bibfnamefont{E.}~\bibnamefont{Bodenschatz}},
  \bibinfo{journal}{Physica D} \textbf{\bibinfo{volume}{193}},
  \bibinfo{pages}{245} (\bibinfo{year}{2004}{\natexlab{a}}).

\bibitem[{\citenamefont{Mordant
  et~al.}(2004{\natexlab{b}})\citenamefont{Mordant, Crawford, and
  Bodenschatz}}]{bib:mordant2004b}
\bibinfo{author}{\bibfnamefont{N.}~\bibnamefont{Mordant}},
  \bibinfo{author}{\bibfnamefont{A.~M.} \bibnamefont{Crawford}},
  \bibnamefont{and}
  \bibinfo{author}{\bibfnamefont{E.}~\bibnamefont{Bodenschatz}},
  \bibinfo{journal}{Physical Review Letters} \textbf{\bibinfo{volume}{93}},
  \bibinfo{pages}{214501} (\bibinfo{year}{2004}{\natexlab{b}}).

\bibitem[{\citenamefont{Monin and Yaglom}(1975)}]{bib:moninbookII}
\bibinfo{author}{\bibfnamefont{A.~S.} \bibnamefont{Monin}} \bibnamefont{and}
  \bibinfo{author}{\bibfnamefont{A.~M.} \bibnamefont{Yaglom}},
  \emph{\bibinfo{title}{Statistical fluid mechanics}},
  vol.~\bibinfo{volume}{II} (\bibinfo{publisher}{MIT press},
  \bibinfo{year}{1975}).

\bibitem[{\citenamefont{Voth et~al.}(2002)\citenamefont{Voth, LaPorta,
  Crawford, Alexander, and Bodenschatz}}]{bib:voth2002}
\bibinfo{author}{\bibfnamefont{G.~A.} \bibnamefont{Voth}},
  \bibinfo{author}{\bibfnamefont{A.}~\bibnamefont{LaPorta}},
  \bibinfo{author}{\bibfnamefont{A.~M.} \bibnamefont{Crawford}},
  \bibinfo{author}{\bibfnamefont{J.}~\bibnamefont{Alexander}},
  \bibnamefont{and}
  \bibinfo{author}{\bibfnamefont{E.}~\bibnamefont{Bodenschatz}},
  \bibinfo{journal}{Journal of Fluid Mechanics} \textbf{\bibinfo{volume}{469}},
  \bibinfo{pages}{121} (\bibinfo{year}{2002}).

\bibitem[{\citenamefont{Ouellette et~al.}(2006)\citenamefont{Ouellette, Xu,
  Bourgoin, and Bodenschatz}}]{bib:Ouellette2006_NJP}
\bibinfo{author}{\bibfnamefont{N.~T.} \bibnamefont{Ouellette}},
  \bibinfo{author}{\bibfnamefont{H.}~\bibnamefont{Xu}},
  \bibinfo{author}{\bibfnamefont{M.}~\bibnamefont{Bourgoin}}, \bibnamefont{and}
  \bibinfo{author}{\bibfnamefont{E.}~\bibnamefont{Bodenschatz}},
  \bibinfo{journal}{NEW JOURNAL OF PHYSICS} \textbf{\bibinfo{volume}{8}},
  \bibinfo{pages}{102} (\bibinfo{year}{2006}).

\bibitem[{\citenamefont{Hill and Wilczak}(1993)}]{bib:hill1993_BullAmPhys}
\bibinfo{author}{\bibfnamefont{R.~J.} \bibnamefont{Hill}} \bibnamefont{and}
  \bibinfo{author}{\bibfnamefont{J.}~\bibnamefont{Wilczak}},
  \bibinfo{journal}{Bulletin of the American Physical Society}
  \textbf{\bibinfo{volume}{38}}, \bibinfo{pages}{2296} (\bibinfo{year}{1993}).

\bibitem[{\citenamefont{Obukhov}(1949)}]{bib:obukhov1949}
\bibinfo{author}{\bibfnamefont{A.~M.} \bibnamefont{Obukhov}},
  \bibinfo{journal}{Doklady Akademii Nauk SSSR} \textbf{\bibinfo{volume}{66}},
  \bibinfo{pages}{17} (\bibinfo{year}{1949}).

\bibitem[{\citenamefont{Gotoh and Fukayanna}(2001)}]{bib:gotoh2001_PRL}
\bibinfo{author}{\bibfnamefont{T.}~\bibnamefont{Gotoh}} \bibnamefont{and}
  \bibinfo{author}{\bibfnamefont{D.}~\bibnamefont{Fukayanna}},
  \bibinfo{journal}{Physical Review Letters} \textbf{\bibinfo{volume}{86}},
  \bibinfo{pages}{3775} (\bibinfo{year}{2001}).

\bibitem[{\citenamefont{Vedula and Yeung}(1999)}]{bib:vedula1999_PoF}
\bibinfo{author}{\bibfnamefont{P.}~\bibnamefont{Vedula}} \bibnamefont{and}
  \bibinfo{author}{\bibfnamefont{P.~K.} \bibnamefont{Yeung}},
  \bibinfo{journal}{Physics of Fluids} \textbf{\bibinfo{volume}{11}},
  \bibinfo{pages}{1208} (\bibinfo{year}{1999}),
  \urlprefix\url{http://link.aip.org/link/?PHF/11/1208/1}.

\bibitem[{\citenamefont{George et~al.}(1984)\citenamefont{George, Beuther, and
  Arndt}}]{bib:george1984_JFM}
\bibinfo{author}{\bibfnamefont{W.~K.} \bibnamefont{George}},
  \bibinfo{author}{\bibfnamefont{P.~D.} \bibnamefont{Beuther}},
  \bibnamefont{and} \bibinfo{author}{\bibfnamefont{R.~E.~A.}
  \bibnamefont{Arndt}}, \bibinfo{journal}{Journal of Fluid Mechanics}
  \textbf{\bibinfo{volume}{148}}, \bibinfo{pages}{155} (\bibinfo{year}{1984}).

\bibitem[{\citenamefont{Nelkin}(1994)}]{bib:nelkin1994_AdvPhys}
\bibinfo{author}{\bibfnamefont{M.}~\bibnamefont{Nelkin}},
  \bibinfo{journal}{Advances in Physics} \textbf{\bibinfo{volume}{43}},
  \bibinfo{pages}{143} (\bibinfo{year}{1994}).

\end{thebibliography}
 
\end{document}